\documentclass[pra,twocolumn,superscriptaddress,showpacs,floatfix]{revtex4}
\usepackage{graphicx,amsmath,amssymb,bm}
\usepackage{amsfonts}

\usepackage{graphicx}
\usepackage{dcolumn}
\usepackage{bm}
\usepackage{braket}

\begin{document}


\title{Toward understanding the microscopic origin of nuclear clustering}


\author{J. Oko{\l}owicz}
\affiliation{Institute of Nuclear Physics, Radzikowskiego 152, PL-31342 Krak\'ow, Poland}%

\author{W. Nazarewicz}
\affiliation{Department of Physics and Astronomy, University of Tennessee, Knoxville, Tennessee 37996, USA
}%
\affiliation{Physics Division, Oak Ridge National Laboratory, Oak Ridge, Tennessee 37831, USA 
}%
\affiliation{Institute of Theoretical Physics, University of Warsaw, ul. Ho\.za 69,
PL-00-681 Warsaw, Poland
}%

\author{M. P{\l}oszajczak}
\affiliation{Grand Acc\'el\'erateur National d'Ions Lourds (GANIL), CEA/DSM - CNRS/IN2P3,
BP 55027, F-14076 Caen Cedex, France
}%

\begin{abstract}
Open Quantum System (OQS) description of a many-body system involves interaction of Shell Model (SM) states through the particle continuum. In realistic nuclear applications, this interaction may lead to collective phenomena in the ensemble of SM states. We claim that the nuclear clustering is an emergent, near-threshold phenomenon, which cannot be elucidated within the Closed Quantum System (CQS) framework. We approach this problem by investigating the near-threshold behavior of Exceptional Points (EPs) in the realistic Continuum Shell Model (CSM).  The consequences for the alpha-clustering phenomenon are discussed.
\end{abstract}

\maketitle

\section{Introduction}
\label{seq1}
The immense richness of the nuclear many-body problem stems  from its genuine multi-scale character and underlying  effective interactions that are strongly mediated by the nuclear medium. Further complexity is added by the OQS nature of 
the atomic nucleus, which requires a treatment of bound states, resonances, and the continuum of scattering states within a unified framework. In standard applications of the nuclear SM, which is a standard tool for nuclear structure studies, the nucleus is described as the CQS where nucleons occupying bound orbits are isolated from the environment of scattering states\cite{rf:1}. 

To formulate the SM for OQSs, two frameworks have been proposed. The first one, the Gamow Shell Model (GSM)\cite{rf:2,rf:3}, is the complex-energy CSM based on the Berggren ensemble\cite{rf:4}. The GSM, which is conveniently formulated in the Rigged Hilbert Space\cite{rf:5}, offers a fully symmetric treatment of bound, resonance, and scattering  states. The second one, the real-energy CSM  in the Hilbert space\cite{rf:6,rf:7,rf:8,rf:9}, is based on the projection formalism\cite{rf:10,rf:11}. A recent realization of this approach, the Shell Model Embedded in the Continuum (SMEC)\cite{rf:7}, provides a unified description of structure and reactions with up to two nucleons in the scattering continuum using realistic SM interactions. The CSM describes many-body states of an OQS using the effective non-Hermitian Hamiltonian in which the coupling to decay channels results in  the anti-Hermitian component. The configuration mixing due to the competition between Hermitian and anti-Hermitian terms is a source of collective features such as, e.g., the resonance trapping\cite{rf:12,rf:13,rf:14} and super-radiance phenomenon\cite{rf:15,rf:16},  multichannel coupling effects in reaction cross-sections\cite{rf:19} and shell occupancies\cite{rf:20},  modification of spectral fluctuations\cite{rf:17}, and deviations from Porter-Thomas resonance widths distribution\cite{rf:14,rf:18}. We shall argue that the appearance of cluster states close to the corresponding cluster emission threshold is yet another consequence of continuum coupling\cite{rf:40}.

Nuclear clustering is arguably one of the most mysterious nuclear phenomena. The comprehensive understanding of its universal occurrence and properties is absent in the CQS formulation of the nuclear many-body problem. The commonly used Cluster Model\cite{rf:21,rf:22} is an {\it a posteriori} approach  that assumes effective building blocks (clusters).  {\it A priori} approaches, like the nuclear SM approach, simply fail to predict cluster states at observed low excitation energies around  cluster-decay thresholds.
In this context, it is worth mentioning that there exists a formal correspondence between Slater determinants of single-particle harmonic oscillator wave functions -- used in many CQS SM frameworks -- and cluster-model wave functions\cite{rf:cmsm}. However, this relation, explicitly utilizing the SU(3) dynamical symmetry of the harmonic oscillator, is of limited use when it comes to quantitative description of cluster states that requires a  proper treatment of decay thresholds and asymptotic behavior of wave functions. In the most advanced approaches to cluster decay, the SM wave functions {\em must be} supplemented 
with a cluster component to provide a quantitative agreement with experiment\cite{rf:cmsm1}. This failure of CQS approaches to describe cluster states  is the central problem in nuclear theory. 

In this study, we shall discuss  collective features of SM eigenstates in the neighborhood of the one-proton decay threshold by analyzing realistic SMEC wave functions. It is expected that genuine features of the  emergent collective behavior of SM states are universal, regardless of the nature of the charged particle (proton, deuteron, alpha-particle, etc.) emission threshold, in analogy with the universal behavior of reaction cross sections\cite{rf:23,rf:19} and spectroscopic quantities\cite{rf:20}. In this sense, the collective behavior of SM eigenstates close to the one-proton decay threshold provides the insight into the formation mechanism of more complicated charged cluster configurations, like alpha-cluster configurations. The SMEC formulation is particularly convenient for our purpose as it allows us to study the collective features of OQSs in terms of the competition of Hermitian (internal) and anti-Hermitian (external,  through the continuum) mixing of SM eigenstates. 

In this work, collective features of CSM (SMEC) states are discussed by analyzing properties of exceptional threads (ETs) of the complex-extended CSM Hamiltonian\cite{rf:24}, i.e., the trajectories of EPs in the space of system energy $E$ and continuum-coupling interaction parameter $V_0$\cite{rf:25}. EPs, corresponding to the entanglement of two sheets of eigenvalues  by the square root  singularity, are results of an interplay between opposite effects of Hermitian and anti-Hermitian parts of the CSM Hamiltonian\cite{rf:8}. The $(E,V_0)$-dependence of EPs  around the proton decay threshold is studied to understand the near-threshold collective rearrangement of SM eigenfunctions, which is at the roots of clustering phenomenon. 

In Sec. \ref{sec2} we shall examine physical arguments that necessitate the use of OQS formalism in the description of nuclear  clustering and recall its universal features. A brief summary of the real-energy CSM and the EPs is presented in Sec. \ref{sec3}. The genuine near-threshold properties of ETs are discussed in Secs.~\ref{sec4} and \ref{sec5} using illustrative examples of SMEC eigenfunctions of $^{16}$Ne and $^{24}$S that are mixed through the continuum coupling to the $\ell=0$ proton decay channel. In particular, we study the role of the so-called aligned state in which the collective strength is concentrated. In Sec. \ref{sec5}, we  investigate the dependence of the continuum-coupling correlation energy and the position of its maximum  on fragment charges and the nuclear radius. These results provide an insight into collective effects in more complex near-threshold situations involving decays of heavier charged particles such as $^3$H, $^2$He, $^4$He, and $^8$Be. Finally, the main results of this work are summarized in Sec. \ref{sec6}.

\section{Configuration mixing near the dissociation threshold}
\label{sec2}
The cluster states are closely related to the nature of a nearby cluster-decay threshold. Ikeda {\it et al.} \cite{rf:26} remarked that alpha-cluster states can be found in the proximity of alpha-particle decay thresholds. Curiously, this insight has not raised much interest in the role of continuum coupling in generating correlations associated with  effective cluster degrees of freedom. 

\begin{figure}[tb]
\begin{center}
\includegraphics[width=0.48\textwidth]{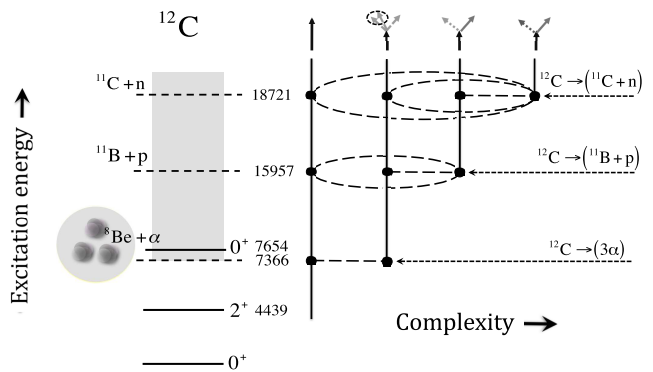}
\end{center} 
\caption{The scheme of couplings in the multichannel representation of $^{12}$C. The structure of a near-threshold Hoyle resonance $0^+_2$ in $^{12}$C exhibits a strong imprint of the nearby  $^{12}$C$\rightarrow ^8$Be+$\alpha$ decay threshold. With the increasing excitation energy, subsequent decay channels open up, leading to an intricate  multichannel network of couplings.}
\label{fig1}
\end{figure}

Actually, the conjecture of Ikeda {\it et al.} can be formulated more generally\cite{rf:40}; namely, the coupling to a nearby particle/cluster decay channel induces particle/cluster correlations in CSM wave functions which are the imprint of this channel. In other words, the clustering is the generic near-threshold phenomenon in OQS which does not originate from any particular property of the nuclear forces or any dynamical symmetry of the nuclear many-body problem. This generalized conjecture holds for all kinds of clustering, including unstable clusters such as dineutron or $^8$Be. Figure \ref{fig1} shows an example of $^{12}$C  for which the lowest-energy threshold corresponds to the emission of an alpha-unbound $^8$Be. 

What can be said about the properties of many-body states  in the narrow range of energies around the reaction threshold? Are they universal, independent of any particular realization of the Hamiltonian? 
The clustering phenomenon can, in fact,  be traced back to basic properties of the scattering matrix in a multichannel network\cite{rf:27}. The decay threshold is a branching point of the scattering matrix. For energies below the lowest particle-decay threshold at energy $E_1$, one finds the analytic phase with a single solution which is regular in the entire space. At energies above $E_1$ and below the next decay channel threshold at energy $E_2$, the new analytic phase corresponds to two regular solutions of the scattering problem, and so on. As a result, one obtains a set of analytic phases, each one with a different number of regular scattering solutions. These phases are separated by decay thresholds and, together, form the multichannel OQS. The coupling of different SM eigenfunctions to the same decay channel modifies regular scattering solutions in the proximity of each branching point. 

Such a network is illustrated schematically in the r.h.s. of Fig.~\ref{fig1}. For energies $E$ below the first decay threshold $E_{^{12}{\text C}\rightarrow 3\alpha}$, the regular phase consists of elastic scattering solutions. For $E>E_{^{12}{\text C}\rightarrow 3\alpha}$ and below the second threshold at energy $E_{^{12}{\text C}\rightarrow ^{11}B+p}$, the regular phase contains two solutions coupled by unitarity, and the story repeats  at higher excitation energies as more and more channels open up. The structure of any eigenfunction in this multichannel OQS depends on the total energy of the system which in turn determines the environment of decay channels accessible for couplings. 
With increasing excitation energy, the multichannel OQS network of couplings becomes more and more complex. The flux conservation (unitarity) implies that the mixing of SM eigenfunctions changes if a new channel becomes available. In this sense, the OQS eigenfunctions are not immutable characteristics of the system but vary with increasing energy of the system\cite{rf:8}.

\section{Continuum Shell Model}
\label{sec3}
The real-energy CSM formulation using Feshbach projection formalism\cite{rf:10,rf:11} offers a convenient framework for a discussion of the competition between  Hermitian and anti-Hermitian components of the effective Hamiltonian. The Hermitian term causes the repulsion of levels in energy and attraction of widths, whereas the anti-Hermitian term leads to 
energy-attraction and width-repulsion effects\cite{rf:8}. The resulting interplay between Hermitian and anti-Hermitian couplings yields a complicated interference pattern\cite{rf:28} and is a source of entanglement of the continuum wave functions\cite{rf:29}. 

The Hermitian mixing of states tends to make decay widths more similar (width equilibration), whereas the anti-Hermitian mixing repels widths leading to the formation of different time scales in the system (width bifurcation). In particular, the anti-Hermitian mixing tends to concentrate the continuum coupling in a single state (aligned state) of the OQS. As discussed below, properties of this state are crucial for the appearance of cluster correlations close to the cluster-decay threshold. 

In the following, we shall employ SMEC to study the  collectivization of SM states around the threshold. 
The detailed description of SMEC can be found elsewhere\cite{rf:8,rf:30} (see also Ref. \cite{rf:16}). The Hilbert space is divided into orthogonal subspaces ${\cal Q}_{\mu}$ where index $\mu$ denotes the number of particles  in the scattering continuum ($\mu=0,1,\dots$). An OQS description of internal dynamics in ${\cal Q}_0$ includes couplings to the environment of decay channels, and is modelled by the energy-dependent  Hamiltonian:
\begin{equation}
{\cal H}(E)=H_0+H_1(E)=H_0+V_0^2h(E) .
\label{eq1}
\end{equation}
In this expression, $H_0$ is the CQS Hamiltonian (the SM Hamiltonian), $V_0$ is the continuum-coupling strength, $E$ is the scattering energy, and $h(E)$ is the coupling term between localized states (${\cal Q}_0$) and the environment of decay channels (${\cal Q}_1$, ${\cal Q}_2$, $\dots$). The `external' mixing of two SM eigenstates $i$ and $j$ due to $H_1(E)$ consists of the Hermitian principal value integral describing virtual excitations to the continuum and the anti-Hermitian residuum that represents the irreversible decay out of the internal space ${\cal Q}_0$. 
The SMEC solutions in ${\cal Q}_0$ are found by solving the eigenproblem for the non-Hermitian Hamiltonian (\ref{eq1}),
\begin{eqnarray}
{\cal H}_{{\cal Q}_0{\cal Q}_0}|\Phi_{j}\rangle&=&{E}_{j}(E,V_0)|\Phi_{j}\rangle \nonumber \\
\langle \Phi_{\bar j}|{\cal H}_{{\cal Q}_0{\cal Q}_0}&=&{E}_{j}^*(E,V_0)\langle \Phi_{\bar j}|,
\label{eqop2}
\end{eqnarray}
in the biorthogonal basis $\langle \Phi_{\bar j}|\Phi_{k}\rangle=\delta_{jk}$. The left $|\Phi_{j}\rangle$ and right $\langle \Phi_{\bar j}|$ eigenstates are related by  complex conjugation. 

An explicit energy dependence of the effective Hamiltonian (\ref{eq1}) is the origin of its strong non-linearity. Moreover, the continuum-coupling term $H_1(E)$ generates effective many-body interactions in the internal space, even if it has  two-body character in the full space. For instance, the continuum-coupling correction to the binding energy induces the change in the effective two-body monopole terms that cannot be distinguished from the effect of a three-body force in SM calculations\cite{rf:31}. 

The  eigenstates $|\Phi_j\rangle$ of ${\cal H}$ are mixtures of SM eigenstates $|\psi_i\rangle$:
\begin{equation}
\label{transf}
|{\psi}_i\rangle \rightarrow |\Phi_j\rangle = {\sum}_{i}^{} b_{ji}|{\psi}_i\rangle,
\end{equation}
where $b_{ji}$ is an orthogonal transformation matrix. The squared matrix element $b_{ji}^2$ is the (complex) weight of the SM eigenstate $\psi_i$ in the correlated OQS eigenstate $|\Phi_j\rangle$:
\begin{equation}
\label{transf1}
{\sum}_i{\cal R}(b_{ji}^2)=1, \hspace{1.5cm} {\sum}_i{\cal I}(b_{ji}^2)=0.
\end{equation}

\begin{figure}[tb]
\begin{center}
\includegraphics[width=0.48\textwidth]{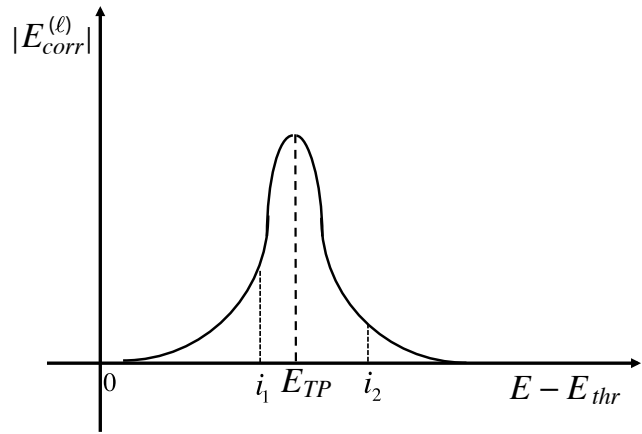}
\end{center} 
\caption{The continuum-coupling energy correction (\ref{eqcorr}) to the SM eigenstates as a function of energy above the proton decay threshold $E_{thr}$. The symbols $i_1$, $i_2$ denote two eigenstates within the energy interval characteristic of a strong coupling to the common decay threshold. The Coulomb interaction suppresses the continuum coupling and shifts the maximum of the correlation energy, $E_{\rm TP}$, above the decay threshold.}
\label{fig2}
\end{figure}

\subsection{Continuum-coupling correlation energy}
\label{sec3.0}

In general, the continuum coupling lowers the binding energies of eigenstates that are close to the decay threshold. The continuum-coupling correlation energy of the SM eigenstate $|\psi_i\rangle$,
\begin{equation}
E_{{\rm corr};i}^{(\ell)}(E) = \langle\psi_i|{\cal H}-H_0|\psi_i\rangle \simeq V_0^2\langle\psi_i|h(E) |\psi_i\rangle, 
\label{eqcorr}
\end{equation}
depends on the structure of the SM eigenstate and  the nature of the decay channel. The correlation energy is peaked  at the threshold only for the coupling to the $\ell=0$ neutron decay channel\cite{rf:42}. For higher partial waves and/or for charged particles, the centrifugal  Coulomb barriers shift the maximum of the correlation energy  above the threshold.

Figure \ref{fig2} shows a typical shape of the continuum-coupling correlation energy (\ref{eqcorr}) for SM states that are coupled to the proton decay channel. The correlation energy is proportional to $V_0^2$ and to the average density of SM states in the energy interval $\varepsilon/2$ on both sides of the decay threshold. An interplay between the Coulomb interaction and the continuum coupling determines the peak of the continuum coupling correlation energy, i.e.,  the full width at half maximum $\sigma$ and the centroid $E_{\rm TP}$  of the correlation energy around the maximum. The maximal value of the continuum coupling energy at the maximum can reach several MeV, depending on the configuration of SM states involved and the nature of the decay channel. Typically, $\sigma$ of the correlation energy peak is less than $\sim$0.5 MeV in case of the coupling to the charge particle decay channel. This is much less than the width of the correlation energy for the coupling to the neutron decay channel, which varies in the range  1 MeV$\leq\sigma\leq$3 MeV\cite{rf:31,rf:42}. The narrowing of $\sigma$ in the charged particle case is a consequence of the Coulomb barrier that cuts off the low-energy tail of the continuum-coupling correlation energy.

Even though the continuum-coupling correlation energy is a small fraction of the total binding energy,  it is of the same order of magnitude as the pairing correlation energy, which profoundly impacts the independent particle motion in the atomic nucleus.  It is therefore pertinent to ask whether correlations induced by the continuum coupling may lead to  instabilities of certain SM eigenstates  near the channel threshold, and whether such couplings are capable of generating collective rearrangements involving many SM eigenstates. This question will be addressed in Secs. \ref{sec4} and \ref{sec5} below. 

\begin{figure}[tb]
\begin{center}
\includegraphics[scale=0.65, angle=00, clip=true]{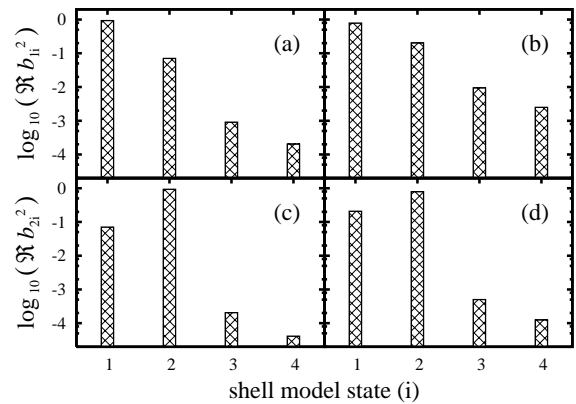}
\end{center} 
\caption{The real part of the weights $b_{ji}^2$ (\ref{transf1}) for $0_1^+$ (top) and $0_2^+$ (bottom) SMEC eigenstates in $^{16}$Ne. In both cases, the OQS eigenstates are at the  maximum of the continuum-coupling correlation energy $E_{\rm TP}$. The continuum-coupling strength is  $V_0=-1$\,GeV fm$^3$ (left) and $V_0=-3$\,GeV fm$^3$ (right).}
\label{fig3}
\end{figure}

An illustration of the continuum-coupling-induced collective features of SM states is the distribution of weights of SM states for a given OQS eigenstate. Figure \ref{fig3} shows real parts of weights $b_{ji}^2$ for the two $0^+$ SMEC eigenstates of $^{16}$Ne. In this calculation, we consider four $J^{\pi}=0^+_i$ SM eigenstates coupled to the $\ell=0$ proton decay channel leading to the  $1/2_1^{+}$  state of $^{15}$F. The SMEC calculations have been carried out  in the  ($0p_{1/2}, 0d_{5/2}, 1s_{1/2}$)  model space, in which there are only four  $0^+$ states in $^{16}$Ne. For $H_0$ we take the ZBM Hamiltonian \cite{rf:32}. The residual coupling between ${\cal Q}_0$ and ${\cal Q}_1$ is generated by the contact force $V_{12}=V_0\delta(r_1-r_2)$. Since the reference energy scale in SM is undefined, we   put the zero energy in SMEC at the lowest particle emission threshold.

The size of the continuum-coupling correlation energy depends on the continuum coupling strength $V_0$, the density of CQS states, and the system excitation energy $E$. The range of relevant $V_0$ values can be determined, e.g., by fitting decay widths of the lowest states in $^{15}$F.  For a ZBM Hamiltonian, the experimental decay widths of the ground state $1/2_1^+$   and the first excited state $5/2_1^+$ in $^{15}$F are reproduced by taking  $V_0=-3500\pm 450$ MeV$\cdot$fm$^3$ and $V_0=-1100\pm 50$ MeV$\cdot$fm$^3$, respectively. The error bars in $V_0$ reflect experimental uncertainties. 

The weights ${\cal R}(b_{1i}^2)$ are shown in Figs.~\ref{fig3}(a) and \ref{fig3}(b) for $V_0=-1$GeV fm$^3$ and  $V_0=-3$GeV fm$^3$, respectively. In this calculation, the lowest eigenstate $\Phi_1$ has an energy $E=E_{\rm TP}$. The largest weight in $\Phi_1$ corresponds to the lowest energy SM eigenstate $\psi_1$. The weights of higher SM eigenstates rapidly drop with $i$. The distribution ${\cal R}(b_{1i}^2)$ becomes broader if the strength of the continuum coupling, or equivalently the density of SM eigenstates, is enhanced. The width of this distribution is a measure of the collective response to the continuum coupling.

The distribution of weights ${\cal R}(b_{2i}^2)$ for the second eigenstate $\Phi_2$ is plotted in Figs.~\ref{fig3}(c) and \ref{fig3}(d). In this case, the eigenstate $\Phi_2$ at $E=E_{\rm TP}$ is close to the threshold and $\Phi_1$ is well bound. Here, the largest weight corresponds to  $\psi_2$. With increasing continuum-coupling strength, the weights of ${\cal R}(b_{22}^2)$ and ${\cal R}(b_{21}^2)$ become comparable. 

\subsection{Double poles of the scattering matrix}
\label{sec3.1}
For $E<0$ (bound system), eigenvalues $E_i(E)$ of the effective Hamiltonian ${\cal H}(E)$ (\ref{eq1}) are real. In the continuum region, $E_i(E)$ correspond to the poles of the scattering matrix and ${\cal H}$ becomes complex-symmetric. The competition between Hermitian and anti-Hermitian parts of ${\cal H}(E)$ may lead to the coalescence of two eigenvalues, i.e., to the formation of  EPs (double poles of the scattering matrix)\cite{rf:25}. The relation of EPs to avoided level crossings and spectral properties of Hermitian systems\cite{rf:33,rf:34} has been discussed in considerable detail in schematic models\cite{rf:35}.

EPs  are indicated by common roots of the two equations:
\begin{eqnarray}
 \frac{\partial^{(\nu)}}{\partial {\cal E}} {\rm det}\left[{\cal H}\left(E;V_0\right)  -{\cal E}I\right] = 0,~~~\nu=0,1.
\label{discr}
\end{eqnarray}
Single-root solutions of Eq. (\ref{discr}) correspond to EPs associated with decaying (capturing) states. For the Hamiltonian (\ref{eq1}), the maximum number of such roots is  
\begin{eqnarray}
M_{max}=2n(n-1),
\label{numroots}
\end{eqnarray}
where $n$ is the number of states of given angular momentum $J$ and parity $\pi$. The factor 2 in Eq.~(\ref{numroots}) comes from the quadratic dependence of ${\cal H}$ on $V_0$ ({\ref{eq1}). 

\begin{figure}[tb]
\begin{center}
\includegraphics[width=0.50\textwidth]{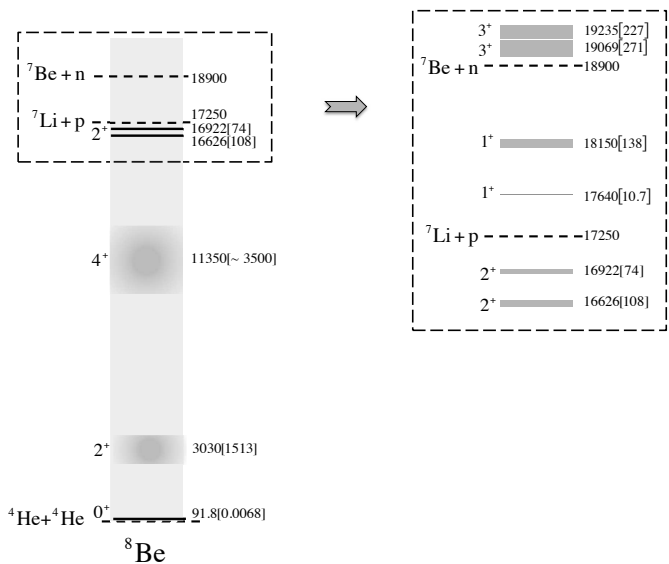}
\end{center} 
\caption{Different phases in the scattering continuum of $^8$Be. Of particular interest are two doublets of nearly degenerate $2^+$ and $3^+$ resonances in the  vicinity of  $^7$Li+p and $^7$Be+n thresholds, respectively.}
\label{fig5}
\end{figure}
An excellent example is the $2^+$ doublet in $^8$Be shown in Fig.~\ref{fig5}, which was extensively discussed in Refs.~\cite{rf:36,rf:37}. These two resonances, which have approximately the structure of $^7$Li+p and $^7$Be+n threshold configurations\cite{rf:38}, are seen below one-nucleon emission thresholds. Above these thresholds, one finds a nearly degenerate doublet of $3^+$ resonances.

The nucleus $^8$Be is an example of how the presence of the reaction threshold can influence the nature of regular solutions. Above the $\alpha$ + $\alpha$ threshold, the ground state $0^+$ and rotational excitations $2^+$, $4^+$ all belong to the the same analytic phase which carries the imprint of the alpha-decay channel. The new phases start slightly below the $^7$Li+p threshold and continue above the $^7$Be+n threshold. The isospin symmetry of the nuclear force and the approximate symmetry of these two configurations with respect to the Coulomb force determine the strong mixing of resonances in these two phases and give rise to the formation of energy doublets that resemble $^7$Li+p and $^7$Be+n mirror thresholds. These $2^+$ and $3^+$ doublets are thus convincing examples of avoided crossing of resonances, the mechanism of which has been explained in a simple two-resonance picture\cite{rf:36}.

The mixing of wave functions in CQSs is closely related to the features of EPs of the complex-extended CQS Hamiltonian. Similarly, to understand the mixing of wave functions in OQSs one should analyze the spectrum of EPs of the complex-extended OQS Hamiltonian. Since the OQS Hamiltonian (\ref{eq1}) is energy dependent, the essential information about configuration mixing is contained in exceptional threads, ETs, the trajectories of coalescing eigenvalues $E_{i_1}(E)=E_{i_2}(E)$  of the effective Hamiltonian for a complex value of  $V_0$\cite{rf:39}. 

\section{Configuration mixing near the charge particle emission threshold}
\label{sec4}
A complete picture of the near-threshold configuration mixing in CSM many-body wave functions can be obtained by investigating properties of ETs of the complex-extended CSM Hamiltonian\cite{rf:40}.  The anti-Hermitian mixing leads to an accumulation of collective effects in a single state close to the particle-emission threshold. Such an aligned state couples strongly to the decay channel and, therefore,  carries many of its characteristics. A large part of the total continuum-coupling correlation energy is concentrated in the aligned state\cite{rf:40} as can be seen in the energy dependence of ETs in Fig. \ref{fig4}.

\begin{figure}[htb]
\begin{center}
\includegraphics[scale=1.15, angle=00, clip=true]{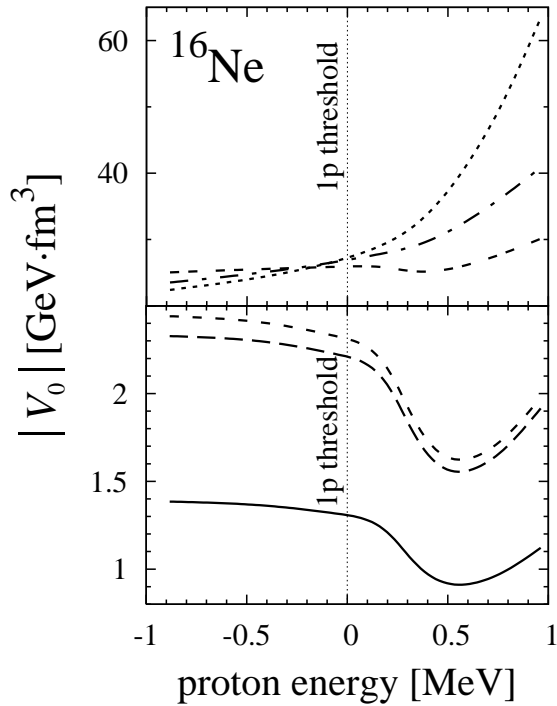}
\end{center} 
\caption{ETs for the $0^+$ eigenvalues of the complex-extended effective Hamiltonian (\ref{eq1}) in $^{16}$Ne plotted as a function of $|V_0|$. Only ETs corresponding to decaying resonances are shown. Each point along an ET is an EP labelled by the proton energy with respect to the 
$^{15}$F$(1/2^+)$+p$(\ell = 0)$ threshold. There are six threads in this case. Three of them (bottom panel), involving the aligned  $0_1^+$ state, are found in the physical range of $|V_0|$ values and correspond to a coalescence of $0_1^+-0_2^+$ (solid line),  $0_1^+-0_3^+$ (long-dashed line), and $0_1^+-0_4^+$ (short-dashed line) eigenvalues. The remaining three threads (top panel), $0_2^+-0_3^+$,  $0_2^+-0_4^+$, and $0_3^+-0_4^+$, correspond to extremely large values of $|V_0|$. For more details of the model space and interaction, see Sec. \ref{sec3}.}
\label{fig4}
\end{figure}
Figure \ref{fig4} displays ETs  for the $0^+$ eigenvalues of the complex-extended SMEC Hamiltonian in 
$^{16}$Ne as a function of  $|V_0|$.  There are four $0^+$ SM eigenstates in this model space and, hence, six different ETs for decaying resonances.
All ETs which are relevant for the collective mixing of SM states involve the aligned eigenstate $0_1^+$. These ETs: $0_1^+-0_2^+$,  $0_1^+-0_3^+$, and $0_1^+-0_4^+$ are shown in the lower panel. They exhibit a minimum of $|V_0|$ (the turning point, TP) at the same proton energy $E_{\rm TP}\simeq0.55$ MeV. At this energy, the collective mixing of SM states is maximal. The remaining ETs, which do not involve the aligned state, are found at exceedingly large values of $|V_0|$ (see the upper panel in Fig.~\ref{fig4}) and have no influence on the configuration mixing.

We assert that the collective mixing of all SM eigenstates with the same quantum numbers $J^\pi$ via the aligned state is the essence of the clustering phenomenon. This mixing, which results from the opening of the many-body system, can explain the appearance of cluster states close to the corresponding cluster decay threshold. The aligned state is also an archetype of the cluster state. 
For charged-cluster configurations, the appearance of clustering depends on whether an SM  state is found inside of the energy window given by the competition between the continuum coupling and the Coulomb interaction (see discussion around Fig.~\ref{fig2}). Consequently, the window of opportunity for charged-particle clustering is situated around the turning point above the charged-cluster decay threshold.

\section{Maximum continuum coupling}
\label{sec5}

The mixing of SM eigenstates through the continuum is strongest in the narrow window of proton energies around the turning point for all ETs which involve the aligned eigenstate. In the following, we shall illustrate some pertinent properties of the turning point. To this end, we consider  $0^+$ SM states in $^{24}$S,  which are coupled to the same one-proton decay channel $^{23}$P($1/2_1^+$) + p($\ell=0$). 
SMEC calculations in $^{24}$S have been  performed in the $(1s_{1/2}, 0d_{5/2}, 0d_{3/2})$ SM model space using the USDB interaction\cite{rf:41}. Since  we consider two $0^+$ states, there are two ETs associated with decaying resonances. 

\begin{figure}[htb]
\begin{center}
\includegraphics[scale=0.66,angle=00,clip=true]{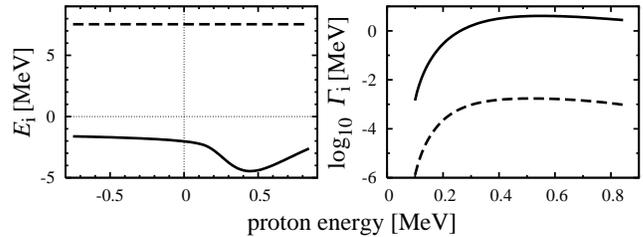}
\end{center} 
\caption{Energies $E_i$ (left) and widths $\Gamma_i$ (right) of $0_i^+$, ($i=1,2$) eigenvalues of the effective Hamiltonian (\ref{eq1}) in $^{24}{\rm S}$   as  a function of the proton energy in the vicinity of the  $^{23}$P($1/2_1^+$) + p($\ell=0$) decay threshold. The aligned state is shown by a solid line. Calculations have been performed using the SMEC Hamiltonian with  $V_0=-1~$GeV~fm$^3$. The threshold energy $E_{thr}$ is put at $E=0$.}
\label{fig6}
\end{figure}
Figure \ref{fig6} shows the energy and width of two $0^+$ eigenvalues of ${\cal H}$ in the vicinity of the $^{23}$P($1/2_1^+$)+p($\ell=0$) decay threshold.   The continuum coupling has the strongest effect on the lowest-energy eigenvalue $E_1$. This aligned eigenstate  becomes fairly broad above the proton emission threshold. There appears a narrow minimum in $E_1(E)$ around the turning point energy $E_{\rm TP}$. In general, $E_{\rm TP}$ depends on the height of the effective (Coulomb+centrifugal)  barrier, the structure of SM wave functions involved, and the decay channel itself. 

\begin{figure}[htb]
\begin{center}
\includegraphics[scale=0.65,angle=00,clip=true]{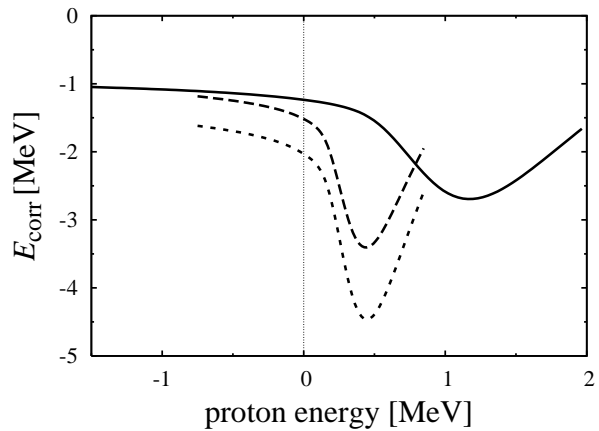}
\end{center} 
\caption{Continuum-coupling correlation energy for the aligned $0^+$ state in $^{24}$S  as  a function of the proton scattering energy for various Coulomb potentials. The solid line corresponds to the Coulomb potential of $^{23}$P+p system ($Z_t\equiv Z_1\cdot Z_2=15$; $Z\equiv Z_1+Z_2=16$) and a uniformly charged sphere of radius $R_0=4$ fm.  The dashed lines correspond to artificially reduced Coulomb potentials with $Z_t=8$ and radii $R_0=3.61$ fm (short-dashed line) and 4 fm (long-dashed line). The latter radius is calculated from the relation $R_0=1.35(A-4)^{1/3}+1.3$ fm and represents the radius of the $^8$Be + $\alpha$ potential ($Z_t=8; Z=6$). Calculations have been performed using the SMEC Hamiltonian with  $V_0=-1~$GeV~fm$^3$.}
\label{fig7}
\end{figure}
Figure \ref{fig7} shows the continuum-coupling correlation energy (\ref{eqcorr}) for the aligned $0^+$  SMEC eigenstate in $^{24}$S, which is calculated for various Coulomb potentials. The solid line is calculated for the Coulomb potential  of  
$^{23}$P+p system ($Z_t\equiv Z_1\cdot Z_2=15$; $Z\equiv Z_1+Z_2=16$, where $Z_1=1$, $Z_2=15$ are proton numbers of the colliding nuclei.). Other curves have been calculated using the same continuum-coupling matrix elements of $^{24}$S and the arbitrarily reduced strength of the Coulomb potential ($Z_t=8$) to assess the impact of the Coulomb potential on the continuum-coupling correlation energy.
It is seen that the correlation energy decreases rapidly with the increasing Coulomb barrier. For the same value of $Z_t$ of fragment charges, the magnitude of the  correlation energy decreases with the increasing radius of the potential $R_0$. Even though the width of the energy window corresponding to large continuum-coupling correlation energy increases with increasing $Z_t$,  it remains significantly smaller than in the neutron case\cite{rf:42}. 

Figure \ref{fig8} shows the energy dependence of the off-diagonal matrix element $b_{12}$ (\ref{transf}) for the $0^+$ aligned state in $^{24}$S.  The minimum of $b_{12}(E)$ corresponds to  the proton energy at which   the largest mixing takes place. This energy is nearly equal to $E_{\rm TP}$. Energy variations of $b_{12}$ with $Z$ and $R_0$ are analogous to those found for $E_{\rm corr}$ (cf. Fig.~\ref{fig7}). 
\begin{figure}[htb]
\begin{center}
\includegraphics[scale=1.15,angle=00,clip=true]{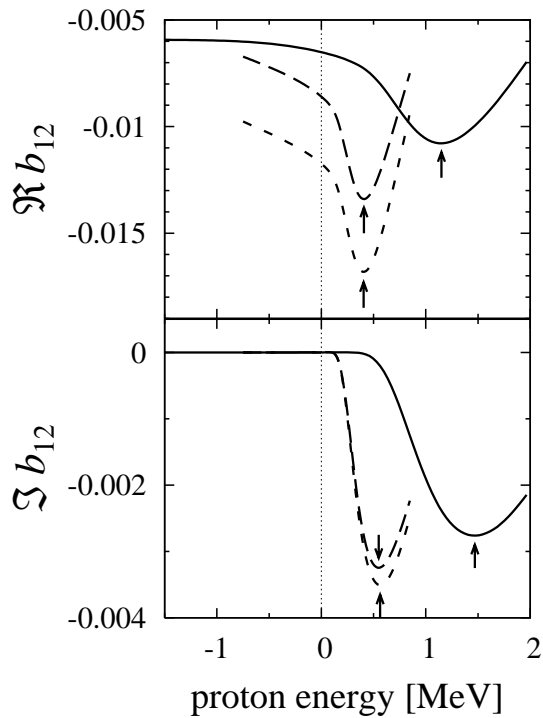}
\end{center} 
\caption{Real and imaginary parts of the amplitudes $b_{ji}$ of the orthogonal transformation (\ref{transf}) for the aligned $0+$ state in $^{24}$S  plotted as  a function of the proton scattering energy for different Coulomb potentials as in Fig.~\ref{fig7}. The point of maximum mixing at $E_{\rm TP}$ is marked by arrows.}
\label{fig8}
\end{figure}
Notice that the magnitude of mixing, as given by $|b_{12}(E)|$, is strongly reduced with increasing proton number $Z$. Consequently, the imprint of the branching point on regular solutions of the scattering problem involving charged particles is reduced in heavier nuclei. This generic feature, with far-reaching consequences for the manifestation of nuclear clustering, does not depend on the particular emission channel and  the angular momentum $\ell$ involved.

The behavior of $E_{\rm corr}(E)$ and $b_{12}(E)$ around the turning point energy $E_{\rm TP}$ follows from properties of ETs of the complex-extended CSM Hamiltonian. Figure \ref{fig10} shows the ETs in the complex-$V_0$ plane calculated for the aligned $0^+$ state in $^{24}$S assuming different  Coulomb barriers.  For each ET, one can see two regions where  the trajectory  turns. The first one, at large values of $|{\cal R}(V_0)|$ and $|{\cal I}(V_0)|$, is seen at the threshold energy $E=E_{thr}$. The second one, at smaller values of $V_0$, is found at the energy $E=E_{\rm TP}$. 
\begin{figure}[htb]
\begin{center}
\includegraphics[scale=0.7,angle=00,clip=true]{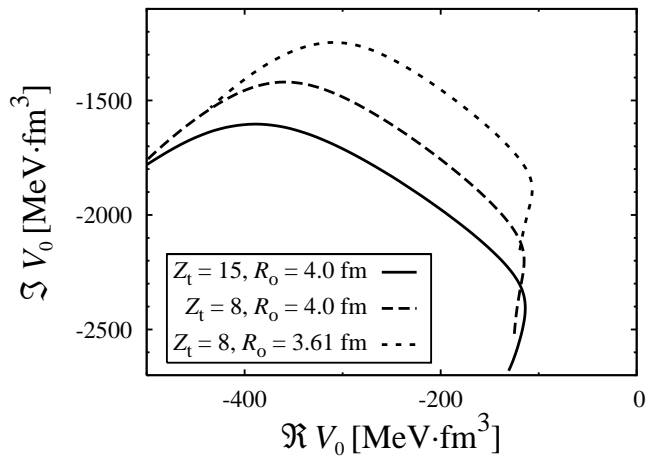}
\end{center} 
\caption{ETs  in the complex-$V_0$ plane for the aligned $0^+$ state in $^{24}$S. SMEC calculations have been performed for  different Coulomb barriers as indicated. The solid line shows results for the standard Coulomb barrier; in the variants marked by dashed lines the Coulomb barrier has been reduced.}
\label{fig10}
\end{figure}
It is obvious from the ${\cal I}(V_0)-{\cal R}(V_0)$ plot that the continuum-coupling induced mixing of SM states diminishes with increasing  $Z_t$  because the turning point at $E=E_{\rm TP}$ moves farther away from the physical limit (${\cal I}V_0$=0) towards larger values of $|V_0|$. This generic tendency does not depend on the nature of the decay threshold. In other words, the increase of the Coulomb barrier in heavier nuclei not only suppresses correlation energy close to proton-decay threshold, but also prevents other kinds of clustering such as the alpha-clustering near the alpha-emission threshold. Conventionally, the latter effect is associated with the disruptive influence of the nuclear spin-orbit interaction on alpha-clustering (see Ref.\cite{rf:43} and references cited therein), i.e., with the specific feature of the nuclear potential.

Whereas, for a fixed value of $Z_t$, the energy $E_{\rm TP}$ of the turning point is independent of the continuum-coupling constant $V_0$, the energy $E_{max}$ corresponding to the maximum continuum correlation energy depends on  $V_0$ that is adjusted to  the experimental data on low-lying resonances. The mismatch between these two energies increases when the `physical' continuum-coupling strength becomes significantly different from $|V_0|(E_{\rm TP})$, i.e., at large values of $Z_t$.

Since $V_0$ determines the overall size of the continuum-coupling term,  the optimal energy window for the appearance of cluster correlations should be largely independent of the detailed structure of  $H_1(E)$ (\ref{eq1}). In other words, given a value of $Z_t$, one expects the turning point energy $E_{\rm TP}$ to be weakly dependent on the nature of the charged particle decay channel and  the parameters of the potential. Only the states in the vicinity of the turning point at $E_{\rm TP}\simeq E_{max}$ may exhibit cluster correlations. Whether they actually do or not depends mainly on a few non-generic parameters such as $V_0$ or the radius of the interaction potential.

\begin{figure}[htb]
\begin{center}
\includegraphics[scale=1.15,angle=00,clip=true]{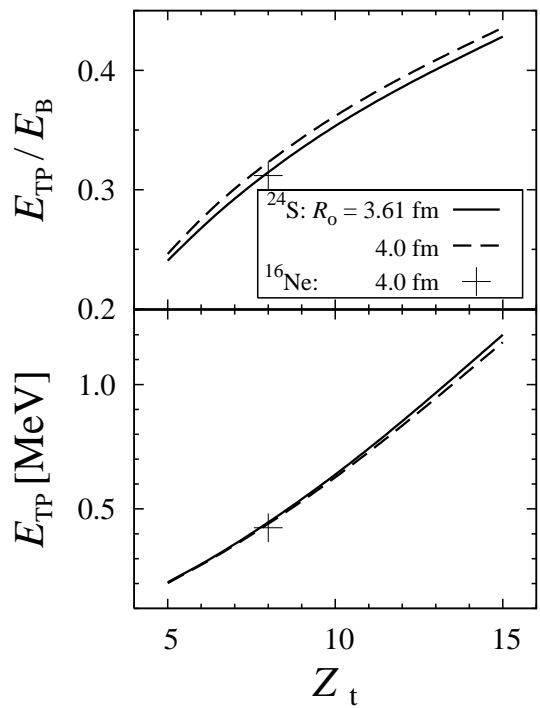}
\end{center} 
\caption{Turning point energy $E_{\rm TP}$ (bottom) and the ratio of $E_{\rm TP}$ to the height of the Coulomb barrier $E_B$ (top) versus $Z_t$ for the aligned $0^+$ states in  $^{24}$S  (lines) and $^{16}$Ne (crosses). SMEC calculations are performed for  $R_0=3.61$\,fm 
(solid line) and 4\,fm (dashed line).}
\label{fig9}
\end{figure}
To see the impact of the Coulomb barrier and the potential radius on $E_{\rm TP}$, the lower panel of  
Fig.~\ref{fig9} shows the dependence of $E_{\rm TP}$ on $Z_t$ in $^{24}$S  calculated in SMEC for two values of $R_0$ and varying arbitrarily the strength of the Coulomb potential. The point $Z_t=8$ and $R_0=4$ fm corresponds to the  $^8$Be + $\alpha$ case, i.e., is relevant to the Hoyle resonance in $^{12}$C. Figure \ref{fig9} also indicates $E_{\rm TP}$ for the aligned state $0^+$ in $^{16}$Ne, which is calculated using $R_0=4$\,fm and the reduced charge $Z_t=8$. This point agrees well with the curve $E_{\rm TP}(Z_t)$ obtained using $^{24}$S  wave functions, i.e.,
$E_{\rm TP}$ seems to be  nearly independent of the potential radius and the detailed structure of SM wave functions involved in the continuum coupling. 
 The upper panel of Fig.~\ref{fig9} displays the ratio of $E_{\rm TP}$ and the height of the Coulomb barrier, $E_B$, as a function of $Z_t$. Also in this representation, the dependence on the potential radius and the detailed structure of SM wave functions are rather weak. Both observations support our assertion that one may use the curve $E_{\rm TP}(Z_t)$ of Fig.~\ref{fig9} to read out the position of the centroid for optimal continuum-induced cluster correlations for any kind of charged clusters. 

\section{Conclusions}
\label{sec6}
Nuclear clustering is a key problem in low-energy nuclear structure studies. The challenging problem is how to separate  generic and specific features of this intricate many-body phenomenon. The energetic order of particle emission thresholds, and their nature, is given by  the $N$- and $Z$-dependence of the nuclear binding energy, i.e., it depends
on specific properties of the nuclear Hamiltonian. The appearance of specific decay channels involving both kinds of nucleons, and  the absence of stable clusters/nuclei entirely composed of like nucleons,  is a direct consequence of the isospin structure of the nuclear force. The continuum coupling effects  play a rather minor role in the interplay  between  various contributions to the total binding energy, at least close to the valley of stability. 

On the other hand, the phenomenological rule that cluster correlations are seen  in the vicinity of the respective cluster emission threshold is unlikely a consequence of specific properties of nuclear forces and calls for a generic explanation. A plausible explanation of this rule, put forward in Ref. \cite{rf:40}, has been in terms of  the collective coupling of SM states via the decay channel(s). This anti-Hermitian coupling leads to the formation of the aligned state, the eigenstate of the OQS which captures  most of the continuum coupling, and, above the decay threshold, exhausts most of the decay width.
 
The mechanism responsible for the creation of an aligned near-threshold state is mathematically similar to the formation mechanism of collective super-radiant or trapped states\cite{rf:12,rf:13,rf:14}. However, the physical domain of  aligned states is not restricted to the region of large density of resonances in the continuum. It can even correspond to  a bound state at energy below the lowest decay threshold. 

The  collectivity of an aligned state is a fingerprint of instability in an ensemble of all SM states having the same quantum numbers and coupled to the same decay channel\cite{rf:40}. Quantitatively,  manifestations of this instability  depend on the strength of the continuum coupling, the density of SM states, and the nature of the decay channel. For large angular momenta $\ell$ and/or charged particle decays, the  energy window for the formation of a collective aligned state is pushed by centrifugal and Coulomb barriers above the decay threshold. 

Are the   examples presented in this study, e.g.,  $J^{\pi}=0^+$ wave functions of $^{24}$S affected by the $^{23}$P($1/2_1^+$) + p($\ell=0)$ proton decay channel, relevant  to  near-threshold clustering effects involving composite clusters?  We have demonstrated  that the point of maximum continuum coupling for the charge-particle decay is related to the turning point of all ETs involving the aligned state. The energy of this turning point is predicted to be fairly {\em independent} on both the continuum-coupling strength and the detailed structure of coupling matrix elements but it varies with the potential radius and the product of fragment charges $Z_t$. This means that one can in principle estimate the position of the energy interval of maximum continuum coupling  for any charge-particle decay channel using the above results for the one-proton decay. On the contrary, the quantitative description of decay properties of the cluster states is more involved and requires a microscopic calculation of both the coupling to the cluster decay channel(s) and the cluster formation amplitude. 

The continuum-coupling correlation energy and the collectivity of mixing amplitudes are reduced  with increasing $Z$ and the charge of emitted particle. This results in  a narrow  energy   window for  the appearance of collective alpha-cluster correlations in medium and heavy nuclei. 

In the studied case of one-proton decay, the aligned state becomes a broad resonance  already at the border of the $sd$-shell.   This problem is less acute for  alpha decay  because of the small preformation factor in question\cite{rf:cmsm1,rf:44}. Consequently,  very narrow aligned alpha-decaying states  can be found around the turning point  in  quite a few $p$- and $sd$- shell nuclei. The Hoyle resonance in $^{12}$C represents  a splendid  example of such a continuum-correlated collective state.
\vskip 0.5truecm

{\bf Acknowledgement}\\
This work has been supported in part by the  MNiSW grant No. N N202 033837, the Collaboration COPIN-GANIL on physics of exotic nuclei, the Project SARFEN (Structure And Reactions For Exotic Nuclei) within the framework of the ERANET NuPNET, FUSTIPEN (French-U.S. Theory Institute for Physics with Exotic Nuclei) under DOE grant number DE-FG02-10ER41700, and by the DOE grant DE-FG02-96ER40963 with the University of Tennessee.

\end{document}